**Integrating Multi -WAN, VPN and IEEE 802.3ad for Advanced IPSEC**

Stefan Ćertić

[Infosec With Experience](#) – Research Department

Information Security

01.02.2024

**Abstract**

Since the emergence of the internet, IPSEC has undergone significant changes due to changes in the type and behavior of users worldwide. IEEE 802.3ad, while considered a key aspect of the IPSEC model, is predictable and can result in potential design flaws, making it relatively easy to access a secure workstation. Thus, it is critical to leverage the benefits of multiple ISPs (multi-WAN) and a link aggregation model and integrate an aspect of randomization in the network. This facet of the network is highlighted by the proof of concept in the simulation of a double pendulum. The analysis of POC provided a network topology designed to utilize multiple WAN, 802.3ad link aggregation, and other environmental components to create a sense of true randomness within a network system. An analysis of this approach shows that it accounts for the data stream's size, transmission speed, WANs and VPNs' location, and other environmental factors to create a sense of randomness. Based on the proof concept, it can be concluded that attaining randomization using multi-WAN, VPN, and 802.3ad is a highly effective model for improving IPSEC.

## Table of Contents

**Integrating Multi -WAN, VPN and IEEE 802.3ad for Advanced IPSEC**

**IPSEC**

## Introduction

Originally born in an academic and military environment, the Internet has become a central component of modern social interaction and business activities. According to Roman et al. (2018), the increased adoption of the Internet in various domains has been predominately beneficial to modern society; these benefits include increased collaboration between groups, companies, and personnel, thereby making innovations and new technology more dynamic in an ever-changing environment (Roman et al., 2018; Zhu et al., 2019). While focusing on the benefits of the internet, the sore subject of trustworthiness emerged; studies (Meng, 2013; Zhu et al., 2019) have shown that online users have undergone tremendous character change from invariably trustworthy to highly alert and suspicious. This change in user attributes has seen the introduction of security-oriented internet protocol (IP) that was not included in the original development and adoption of the internet.

The reduced trust among internet users, while critical to the introduction of internet protocol security, did not occur randomly at once. It was subjected to a series of considerations and progressive analysis to determine the uses and overall impacts on information transfer (Aras et al., 2019). This facet saw the process become continuous, as shown by the different eras of network security; for instance, the packet filtering era was more concerned with malicious hacking and the protection of computer systems from other forms of malware attacks. The session inspection era was more focused on the communication process to determine the legitimacy of data exchange, while the modern era (application control era) was more oriented towards unified threat management across all devices. The dynamism of these areas saw the contentious experimentation of different security protocols with the aim of attaining the most secure data transmission protocol. Despite the different efforts, there has been minimal discussion on whether the integration of multiple WANs and 802.3ad would significantly advance the existing transmission control protocol (TCP).

## Existing Efforts to Improve Modern IPSEC

Today, technological systems are increasingly becoming smaller as their CPU power increases; this increase has been accompanied by capabilities, especially in terms of statistical analysis. At the same time, Zhu et al. (2019) recognized the presence of distributed computing clusters that have been equipped with various CPU nodes and higher RAM powered for faster decryption of security keys. As a result, a study by Li et al. (2016) recognized the presence of a multipath transmission system that is highly beneficial to internet protocol security; specifically, smart devices are known to have multiple network interfaces that have resulted in the adoption of tunnelling mechanisms for better connectivity. Despite the advantages associated with this solution, it is essential to recognize that this approach is hidden from the transport protocols. The methodology also allows an application-level connectivity for a more distributed access to the data packets over a given network, regardless of the location— with the increased processing and analytical powers of modern computing, the approach present significant risks to various networks, especially if the approach is utilized as part of the internet protocol security.

The introduction of the multipath transmission in IP security, especially in Transmission control protocols (TCP), highlights the continuous efforts to changes and improve not only internet security but also the entire the technological landscape in terms of processing power and potential to advance IP Security (IPSEC). However, not all changes in IP security can be realized due to the different risks they present and the underlying scalability challenges. As a result, Li et al. (2016) opted to create an approach that accounts for variation in metrics in the modern age, thereby creating room for a scalable multipath approach to improving IPSEC. According to Li et al. (2016), the highly embraced Internet of Things has created a new vision for network security through the utilization of low-power wide area networks (LPWAN). An analysis of the approach shows that the system offers multi-km communication at data rates of between 0.006 kbps and approximately 10 mbps. Li et al. (2016) found this approach to be highly effective if cryptographic frequency hooting is used to mitigate against jamming— the results were subject to timing, the spread factor, and frequency selection as a way of allowing multiple unconfirmed communications to be undertaken simultaneously with minimal contention.

Li et al.'s (2016) analysis of internet protocol security shows the ability of modern technology to evolve and account for various needs associated with the Internet of Things (IoT) and data security. These insights, while critical, utilize different techniques that can best be described as predictable, thus, can be overridden, posing further security risks. Based on this weakness, it is critical to explore the other WAN-based techniques that rely on randomness for a secure TCP— accounting for the inclusion of the 802.3ad should also be central to the adoption.

**Multiple WAN and 802.3ad for advanced IP security**

An assessment of modern network systems shows an increased emphasis on performance and secure communication and data transfer. This approach to networking, while critical, has seen the adoption of inconsistent and redundant protocols such as IKEv1, creating a need for a more simplified yet secure approach to IPSEC (Jiang et al., 2019). This need saw the release of the IKEv2, which simplified the complexities associated with IKEv2 and provided security on multiple entry points, making it highly dynamic to be used in a wide area network. While focusing on these improvements, Jiang et al. (2019) recognized that threats are increasingly becoming more advanced, hence the need for a more dynamic approach to internet security; for this reason, the adoption of multiple WAN accompanied by VPN has been proposed by multiple studies (Tongkaw & Tongkaw, 2018; Jiang et al., 2019). The adoption of multiple WANs and VPNs in the proposed IPSEC systems was designed to eliminate the overreliance on the AES, SA, and sha256 algorithms by integration of the SM2-SM3 certificates for more dense storage and higher level encryption than existing Internet Protocol security measures.

The adoption of new certificates and encryption models it set to allow integration of new security protocols in the underlying security mechanism; despite such a unique approach, it is vital to acknowledge the potential for human factor and other external occurrences to shaping the overall dynamics of the security protocols. Specifically, the presence of human factors implies that decryption TCP may be far easier due to individual specific preferences, thereby creating a potential weak endpoint in the network. To account for this flaw, the proposed network model integrates link aggregation protocols, the IEEE 802.3ad, as a mechanism to bide the different ISPs and VPNs in the form of a single link (Lacković & Tomić, 2017). An analysis of the aggregated link shows that it contrails the bandwidth of all links within the network such that failure in a single group of interfaces, traffic is rerouted to other available interfaces— this allows continuity in network performance, with the capacity to identify the underlying problems through the nautical reduction in bandwidth. Given these dynamics of the network, the final design of the proposed IPSEC is shown in the figure below:

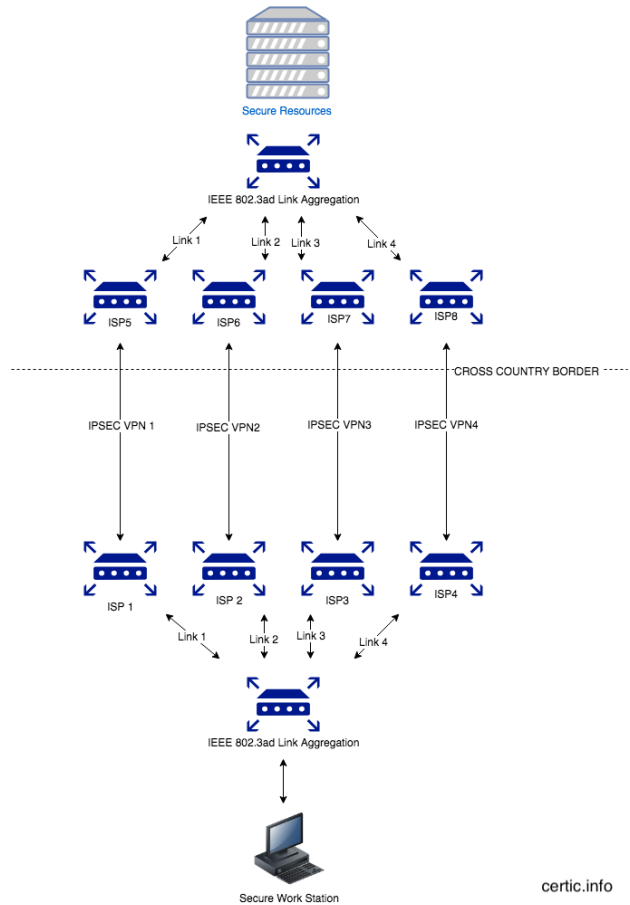

The introduction of IEEE 802.3ad, while considered a great addition to the proposed IPSEC model, still has an element of predictability that may result in potential design flaws, making it relatively easy to access a secure workstation. For this reason, it was vital to leverage the benefits of multiple ISPs (multi-WAN) and a link aggregation model and integrate an element of randomization in the network. The dynamics of this inclusion are such that it uses the link between the ISPs and different endpoints as vectors of connected bodies, with constant factors governing the motion of the said vectors. This facet of the network is highlighted by the proof of concept in Elbori & Abdalsmd's (2017) simulation of a double pendulum.

**Proof of Concept and Discussion**

Based on specifications of the design and functionality presented in Figure 1, consider data traffic from one endpoint to another— this can equated to a pendulum, that is, it can move freely from one WLAN to another through specified routes with a constant factor that creates an equilibrium. In a double pendulum system, it is often convenient to determine coordinates as a factor of angles between the rods and the vertical. As such, let $m_1, L_1 \ and \ \theta_1$ be the mass, length, and angle from the normal of the inner bob, as shown in Figure 2 below— similarly, let the mass, length, and angle from the normal of the inner bob be repressed as $m_2, L_2, and \ \theta_2$.

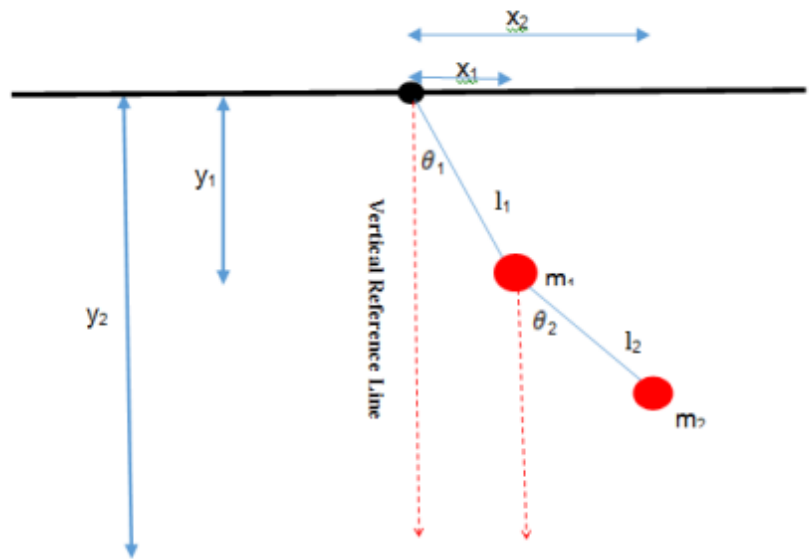

The variables highlighted in Figure 1 above provide the position of the bobs in relation to the normal; as such, the position of the outer bob is given by $(x_1, y_1)$, while that of the inner bob is given as $(x_2, x_2)$. While the static position of the nodes is vital, the option of someone swinging the pendulum in a specific direction needs to be accounted for. For this reason, the kinematics of the pendulum based on the coordinates can be divided using Lagrange's equation.

When adopting Lagrange's equation, Elbori and Abdalsmd (2017) recommend focusing on ensuring the simplicity of numerical analysis. Through this consideration, the first goal is to account for the kinetics if the mass of the inner and outer bob are equal and their lengths are equal ($m_1 = m_2$ and $L_1 = L_2$). In such a scenario: $m_1 = m_2 = m$ and $L_1 = L_2 = l$. This is because we consider the presence of two identical rods, where $f(I)$ can be represented as a product of the mass and length of the pendulum; hence $I = \frac{1}{12}ml^2$. Given the $f(I)$, assume that the mass of the rods is negligible, but their moment of inertia must be included for a better representation of the physical system. As such, obtaining the actual position of the bobs will require accounting for the trigonometric equations associated with the setup, as shown in Figure 1; therefore, the following positional equations can be obtained:

$$x_1 = \frac{1}{2}\sin(\theta_1)$$

For $x_2$, it is vital to account for both. $\theta_1$ and $\theta_2$; therefore, the position of $x_2$ is obtained by $l(\sin(\theta_1)) + \frac{1}{2}\sin(\theta_2)$. For $y_1$, the equation $-\frac{1}{2}\cos(\theta_1)$ is used; in the case of $y_2$ the equation $-l(\cos(\theta_1)) + \frac{1}{2}\cos(\theta_2)$ is used.

Once the position of the bobs has been obtained, the Lagrange equation (L = Kinetic Energy – Potential Energy) can be introduced to determine the first term linear kinetic energy. However, data transmission over a network occurs at different bandwidths and tends to be of different sizes— this created the need to account for the scenarios where $m_1 \neq m_2$ and $L_1 \neq L_2$. Subsequently, the focus shifted from accounting for the overall kinematics of the system to focusing on the momentum and behavior of the system. Depending on the initial coordinates, Elbori & Abdalsmd's (2017) analysis shows that the resulting movement may be periodic, quasi-periodic, or chaotic, creating a sense of true randomness in the interaction between various WAN, VPNs, and components within the network. An aggregation link allows for expansive encryption by introducing different environmental factors that are not random by nature, making statistical analysis impossible.

## Conclusion

Since the introduction of the Internet, Internet Protocol security (IPSEC) has undergone tremendous changes due to changes in the type and behavior of users across the globe. An analysis of these changes shows increased emphasis on the encryption of TCP for secure communication between servers, networks, and workstations. While these facets of TCP have been crucial to reducing security issues associated with internet protocols and cybercrime, they have not been sufficient security to ensure that no encryption is cracked or decrypted. As such, the analysis provided a network topology designed to utilize multiple WAN, 802.3ad link aggregation, and other environmental components to create a sense of true randomness within a network system. An analysis of this approach shows that it accounts for the size of the data stream, speed of transmission, location of WANs and VPNs, as well as other environmental factors to create a sense of randomness. This facet is best highlighted by the simulation of a double pendulum whose motion largely depends on the masses of the bobs, length of the rods, and initial position— it has a tendency to create diverse motions, resulting in highly chaotic patterns, especially when other environmental factors are present. Based on the proof concept and studies conducted on the subject, one can conclude that going truly random using multi-WAN, VPN, and 802.3ad is a highly effective model for improving IPSEC.